\documentclass{article}

\usepackage{microtype}
\usepackage{graphicx}
\usepackage{subcaption}
\usepackage{booktabs} 
\usepackage{multirow}

\usepackage{hyperref}

\usepackage[accepted]{icml2026}

\usepackage{amsmath}
\usepackage{amssymb}
\usepackage{mathtools}
\usepackage{amsthm}

\usepackage[capitalize,noabbrev]{cleveref}

\theoremstyle{plain}

\theoremstyle{definition}

\theoremstyle{remark}

\usepackage[textsize=tiny]{todonotes}

\begin{document}

\twocolumn[
    \icmltitle{Equivariant Graph Neural Networks Improve Optical Spectra Prediction for Materials Screening}
    \icmltitlerunning{Equivariant Graph Neural Networks Improve Optical Spectra Prediction for Materials Screening}



    \icmlsetsymbol{equal}{*}

    \begin{icmlauthorlist}
        \icmlauthor{Kasper Helverskov Petersen}{yyy}
        \icmlauthor{François Raymond J Cornet}{yyy}
        \icmlauthor{Martin Ovesen}{xxx}
        \icmlauthor{Mikkel Jordahn}{yyy}
        \icmlauthor{Kristian S. Thygesen}{xxx}
        \icmlauthor{Mikkel N. Schmidt}{yyy}
    \end{icmlauthorlist}

    \icmlaffiliation{yyy}{Department of Applied Mathematics and Computer Science, Technical University of Denmark, Kongens Lyngby, Denmark}
    \icmlaffiliation{xxx}{Department of Physics, Technical University of Denmark, Kongens Lyngby, Denmark}

    \icmlcorrespondingauthor{Kasper Helverskov Petersen}{khepe@dtu.dk}

    \icmlkeywords{Machine Learning, ICML}

    \vskip 0.3in
]



\printAffiliationsAndNotice{}  

\begin{abstract}
    Scalable prediction of optical spectra is a critical component of high-throughput materials screening for optoelectronic applications such as solar cells.
    Existing surrogate models are trained on spectra computed from lower levels of theory or rely on rotation-invariant scalar features, limiting their geometric expressiveness.
    We explore the use of equivariant graph neural networks for optical spectra prediction, adapting GotenNet to this task and evaluating it on multiple datasets including a recently published collection of 10,533 structures with spectra computed at the level of the random phase approximation (RPA). The proposed model outperforms the current state of the art, with the largest gains in the 0--8\,eV range and on predicting the static real permittivity, both of particular relevance for thin-film optics.
\end{abstract}

\section{Introduction}
AI is increasingly applied across scientific disciplines, and materials discovery, sitting at the confluence of physics, chemistry, and engineering, is a natural target. The combinatorial vastness of chemical and structural space makes AI-driven exploration particularly compelling: generative models propose candidate materials; surrogate models screen them for target properties; and first-principles calculations and experimental characterization evaluate the most promising candidates. Together, these form a pipeline bridging machine learning, computational physics, and materials synthesis and characterization. 
For discovering improved materials for solar cells, a critical component is the scalable prediction of optical spectra. Experimental measurement and ab initio methods such as density functional theory (DFT)~\cite{dft,kohn-sham} and many-body perturbation theory (MBPT)~\cite{mbpt} are resource-intensive and computationally prohibitive for large-scale screening. Machine learning surrogates trained on first-principles databases offer a compelling alternative, but must be accurate enough to reliably distinguish promising candidates from poor ones, making architectural improvements a key priority.

Existing approaches~\cite{GNNOpt, ibrahim2024, alignn} are trained on limited data at lower levels of theory, such as the independent-particle approximation (IPA) or tight-binding (TB). Meanwhile, \citet{OptiMate3B} recently introduced a dataset of 10,533 optical spectra at the level of the random phase approximation (RPA) using an IPA-pretrained model and achieving state-of-the-art results. Their architecture, OptiMate3B, encodes geometry only through rotation-invariant scalar features. Furthermore, it explicitly constructs a line-graph over edges to capture angular features, which necessitates a Voronoi graph construction to limit connectivity. Equivariant architectures avoid both limitations: they encode directional information by construction, providing a geometric inductive bias that is especially valuable given limited training data, and capture angular correlations implicitly, allowing distance-based graph construction that more naturally reflects the physical interaction ranges of atoms.

In this work, we adapt GotenNet~\cite{GotenNet}, a state-of-the-art equivariant graph neural network (GNN), to optical spectra prediction, and evaluate it on both the IPA and RPA OptiMate3B datasets. The resulting model, \(\text{GotenNet}_\text{Opt}\), achieves consistent improvements with the most pronounced gains in the 0--8 eV range and on predicting the static real permittivity, both of particular relevance for materials screening in thin-film optics. We further benchmark against GNNOpt~\cite{GNNOpt} on their smaller dataset of 944 IPA-computed spectra, demonstrating improvements in a low-data setting.


\section{Background}
\subsection{The dielectric function}
\label{sec:dielectric_function}
The predicted optical spectra consist of the complex dielectric function, which describes the response of a solid to an applied electric field $\mathbf{E}$ of angular frequency $\omega$. The dielectric function is actually a second-rank tensor, $\varepsilon_{ij}(\omega)$, relating the electric polarization density along the Cartesian direction $i$ induced by an electric field polarized along the direction $j$. The tensorial nature reflects that the polarization
response can differ along different crystallographic directions. Mathematically, this can be expressed as the linear relationship between the components of the electric displacement field $\mathbf{D}$ and the components of the applied electric field $\mathbf{E}$
\begin{equation}
    D_i(\omega) = \varepsilon_0\sum_j\varepsilon_{ij}(\omega) E_j(\omega),
\end{equation}
where $\varepsilon_0$ is the permittivity of free space and the second-rank tensor is unitless. Alternatively, the second-rank tensor can be formulated as the electric susceptibility, $\chi_{ij}$, describing the linear relationship between the polarization density $\mathbf{P}$ and the applied electric field
\begin{equation}
    P_i(\omega)=\varepsilon_0\sum_j\chi_{ij}(\omega)E_j(\omega),
\end{equation}
where the relationship between $\varepsilon_{ij}$ and $\chi_{ij}$ is simply
\begin{equation}
    \varepsilon_{ij}(\omega)=\delta_{ij}+\chi_{ij}(\omega).
\end{equation}
Typically, determining all 9 tensor elements is unnecessary due to the crystal structure being invariant w.r.t.~certain sets of symmetry operations. For example, for materials with cubic symmetry or isotropic structure, the off-diagonal elements of the tensor vanish while the diagonal elements reduce to a scalar, i.e. $\varepsilon_{ij}(\omega) = \varepsilon(\omega)\delta_{ij}$. More generally, the isotropic part of the dielectric function tensor can be defined as
\begin{equation}
    \bar{\varepsilon}(\omega) = \text{Tr}(\varepsilon(\omega))/3,
\end{equation}
which captures the mean linear response across all crystallographic directions. It consists of a real and imaginary part which are related to the refractive index $n(\omega)$ and the extinction coefficient $\kappa(\omega)$, describing the amount of light attenuation, through
\begin{equation}
    n(\omega)+i\kappa(\omega) = \sqrt{\bar{\varepsilon}(\omega)},
\end{equation}
which can also be written as
\begin{equation}
    n=\sqrt{\frac{\sqrt{\text{Re}(\bar{\varepsilon})^2+\text{Im}(\bar{\varepsilon})^2}+\text{Re}(\bar{\varepsilon})}{2}},
\end{equation}
and
\begin{equation}
    \kappa=\sqrt{\frac{\sqrt{\text{Re}(\bar{\varepsilon})^2+\text{Im}(\bar{\varepsilon})^2}-\text{Re}(\bar{\varepsilon})}{2}}.
\end{equation}
While the real and imaginary parts of $\bar{\varepsilon}$ (or equivalently, $\bar{\chi}$) can be calculated independently of each other, a consequence of causality is that they must satisfy the criterion that they are related to each other by Hilbert transformations~\cite{toll1956causality}
\begin{equation}
    \text{Re}\left(\bar{\chi}(\omega)\right)=\frac{1}{\pi}\mathcal{P}\int_{-\infty}^{\infty}\frac{\text{Im}\left(\bar{\chi}\left(\omega'\right)\right)}{\omega'-\omega}\text{d}\omega',
    \label{eq:kramerskronig1}
\end{equation}
and
\begin{equation}
    \text{Im}\left(\bar{\chi}(\omega)\right)=-\frac{1}{\pi}\mathcal{P}\int_{-\infty}^{\infty}\frac{\text{Re}\left(\bar{\chi}\left(\omega'\right)\right)}{\omega'-\omega}\text{d}\omega',
    \label{eq:kramerskronig2}
\end{equation}
where $\mathcal{P}$ denotes that the singularity at $\omega'=\omega$ is resolved as the Cauchy principal value. The equations \eqref{eq:kramerskronig1} and \eqref{eq:kramerskronig2} are known as the Kramers-Kronig relations.\par\bigskip
For the OptiMate3B dataset, we predict the concatenated real and imaginary parts of $\bar{\varepsilon}(\omega)$, sampled from 0 to 20\,eV in 10\,meV steps, yielding
\begin{equation}
    \text{out} = \text{Re}[\text{Tr}(\varepsilon(\omega))/3] \, \| \,
    \text{Im}[\text{Tr}(\varepsilon(\omega))/3],
\end{equation}
which is a 4002-dimensional output vector. Whereas, for the GNNOpt dataset, a separate model is trained and evaluated for the real and imaginary components, with each predicting a single component interpolated onto a uniform grid of 251 points over $0 \leq \hbar\omega \leq 50$\,eV, where $\hbar$ is the reduced Planck constant:
\begin{equation}
    \text{out}_{\text{Re}} = \text{Re}[\text{Tr}(\varepsilon(\omega))/3], \qquad
    \text{out}_{\text{Im}} = \text{Im}[\text{Tr}(\varepsilon(\omega))/3],
\end{equation}
each yielding a 251-dimensional output vector. A brief background section on DFT is provided in~\cref{app:dft_background}.

\subsection{3D graph representation of crystal structures}
A crystal is an infinite periodic solid described by a unit cell: a parallelepiped spanned by lattice vectors $\mathbf{a}_1, \mathbf{a}_2, \mathbf{a}_3 \in \mathbb{R}^3$ and a finite set of atoms at positions $\{\mathbf{x}_i\}$, given in Cartesian or fractional coordinates $\{\mathbf{s}_i\} \subset [0,1)^3$ via $\mathbf{x}_i = \sum_k s_{i,k}\mathbf{a}_k$. Under periodic boundary conditions (PBC), each atom at $\mathbf{x}_i$ has images at $\mathbf{x}_i + \sum_k n_k \mathbf{a}_k$ for all $n_k \in \mathbb{Z}$.

We represent crystals as a graph $G = (V, E)$, where nodes $V$ are the unit cell atoms and edges $E$ encode atomic interactions, with the neighborhood of atom $i$ defined as $\mathcal{N}(i) = \{ j \in V \mid (i, j) \in E \}$. PBC means the neighborhood search includes periodic images, so multiple edges may exist between the same pair of atoms, each with a distinct edge vector $\mathbf{r}_{ij}$. We use a distance-based cutoff of $r_c = 6.0$\,\AA, and additionally evaluate a Voronoi-based construction, which yields fewer edges and is the method used by OptiMate3B.

Each atom $i$ is featurized by concatenating one-hot encodings of its periodic table group (18 classes) and period (7 classes), giving $\mathbf{z}_i \in \mathbb{R}^{25}$. Each edge carries the unit vector $\hat{\mathbf{r}}_{ij} = \mathbf{r}_{ij}/\|\mathbf{r}_{ij}\|$ and the interatomic distance $\|\mathbf{r}_{ij}\|$.

\subsection{Equivariant graph neural networks}
GNN architectures for atomic systems are commonly formalized within the message passing neural network (MPNN) framework~\cite{gilmer}, where atom embeddings are iteratively updated by aggregating information from neighbors. At each round $t \in \{1, \dots, T\}$, messages are computed and aggregated, the embeddings updated, and after $T$ rounds a readout layer produces the final output:
\begin{align}
    \mathbf{m}_i^{(t+1)} & = \bigoplus_{j \in \mathcal{N}(i)} M_t\!\left(\mathbf{h}_i^{(t)}, \mathbf{h}_j^{(t)}\right), \\
    \mathbf{h}_i^{(t+1)} & = U_t\!\left(\mathbf{h}_i^{(t)}, \mathbf{m}_i^{(t+1)}\right), \\
    \hat{y}              & = R\!\left(\left\{\mathbf{h}_i^{(T)} \mid i \in V\right\}\right),
\end{align}
where $M_t$, $U_t$, and $R$ are the message, update, and readout functions, and $\bigoplus$ is a permutation-invariant aggregation.

Early architectures relied on invariant scalar features derived from pairwise distances, angles, and torsions~\cite{gilmer, schnet, mgcn, dimenet, spherenet}. Equivariant architectures instead propagate vector or higher-order tensor features that transform consistently under symmetry operations~\cite{segnn, egnn, painn, eqgat, torchmdnet, mace, tensornet, equiformer, GotenNet}. Formally, a function $f$ is equivariant to the action of a group $\mathcal{G}$ if
\begin{equation}
    f(g \cdot \mathbf{x}) = g \cdot f(\mathbf{x}) \quad \forall\, g \in \mathcal{G}.
\end{equation}
By design, this property guarantees orientation-consistent predictions, eliminating the need for the model to learn these geometric symmetries directly from the training data.

\section{Method}

\subsection{\(\text{GotenNet}_\text{Opt}\) Architecture}
\begin{figure*}[t]
    \centering
    \includegraphics[width=0.9\textwidth]{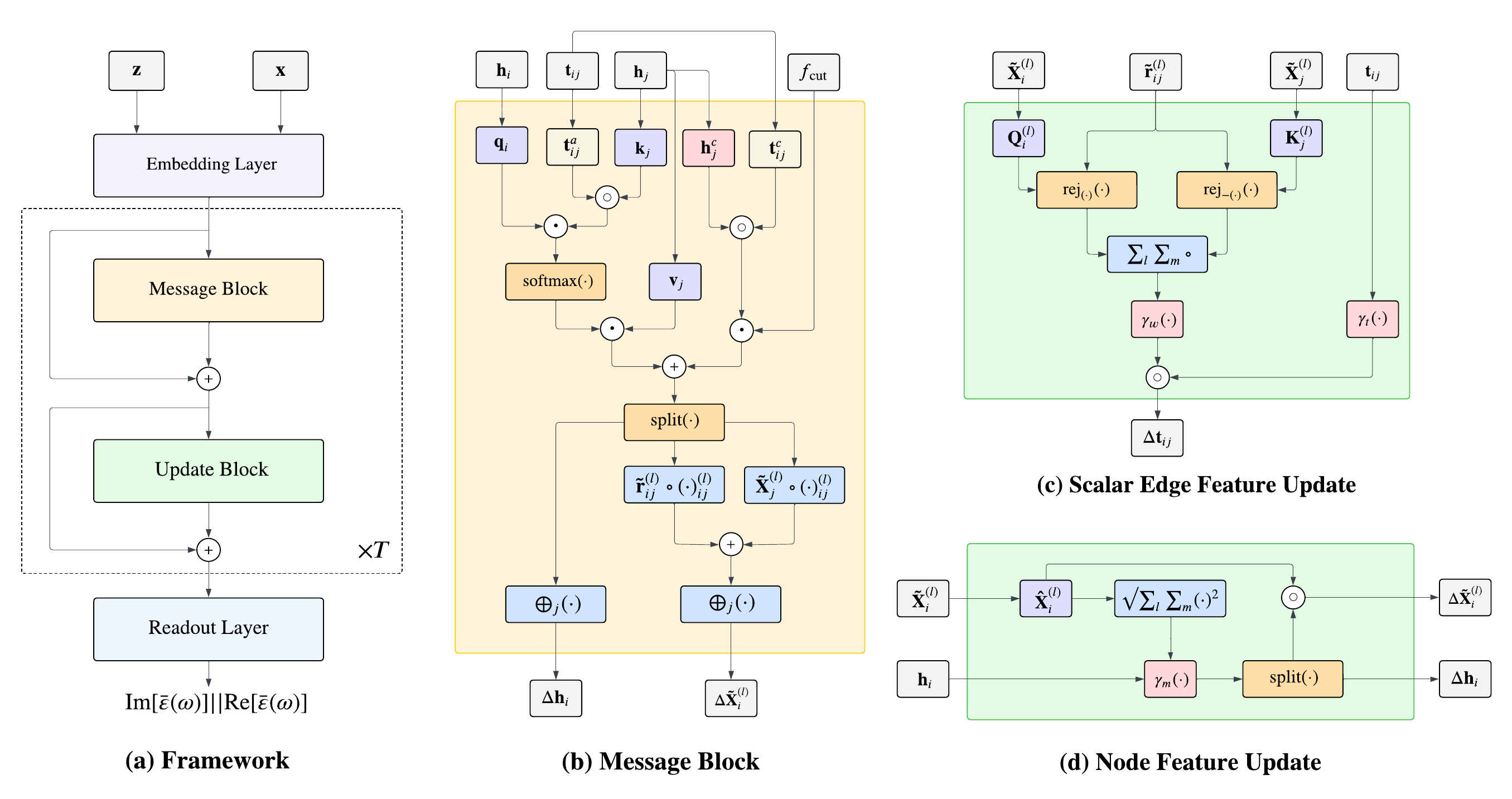}
    \caption{Architecture of \(\text{GotenNet}_\text{Opt}\). (a) The model follows the MPNN framework: an embedding layer initializes atomic and bond features from elemental information and 3D coordinates, \(T\) message-passing rounds update the representations, and a readout layer predicts the optical spectra. (b) The message block. (c) Scalar edge feature update. (d) Node feature update. The summation \(\sum_l\sum_m\) is over \(l=1,\dots,L_\text{max}\) and \(m=2l+1\). Symbols: \(\cdot\) dot product, \(\circ\) element-wise product. Figure adapted from~\cite{GotenNet}.}
    \label{fig:architecture}
\end{figure*}

\cref{fig:architecture} outlines the proposed architecture, which builds upon GotenNet~\cite{GotenNet}. For clarity, we reformulate the model below and highlight our specific modifications.

\subsubsection{Tensor Features}
For each edge $(i,j)$, we encode tensor edge features:
$\tilde{\mathbf{r}}_{ij} = \{\tilde{\mathbf{r}}_{ij}^{(l)}\}_{l=0}^{L_{\max}}$, where $\tilde{\mathbf{r}}_{ij}^{(0)} = \|\mathbf{r}_{ij}\|$ is the scalar distance, $\tilde{\mathbf{r}}_{ij}^{(1)} = \hat{\mathbf{r}}_{ij}$ is the normalized direction vector, and $\tilde{\mathbf{r}}_{ij}^{(l)} = Y^{(l)}(\hat{\mathbf{r}}_{ij}) \in \mathbb{R}^{2l+1}$ for $l>1$ are degree-$l$ real spherical harmonics in Cartesian coordinates, normalized to unit norm. Each node $i$ similarly carries steerable features $\{\tilde{\mathbf{X}}_i^{(l)}\}_{l=1}^{L_{\max}}$, with $\tilde{\mathbf{X}}_i^{(l)} \in \mathbb{R}^{(2l+1)\times F}$ transforming as a rank-$l$ irreducible representation of SO(3). Both the edge and node tensor features thus capture increasingly higher-order geometric information: degree $l=1$ encodes directional information, while higher degrees encode more complex angular patterns such as planar orientations ($l=2$). The edge tensor features are fixed geometric inputs, while the steerable node features are initialized to zero and iteratively updated across message passing layers.

\subsubsection{Embedding Layer}
\paragraph{Scalar node feature initialization.}
Scalar node features $\mathbf{h}_{i} \in \mathbb{R}^{F}$ are initialized by aggregating over neighbors $j \in \mathcal{N}(i)$ using element-wise products of atom, covalent radius, and distance embeddings.
Inspired by GNNOpt~\cite{GNNOpt}, which demonstrated that incorporating covalent radius as an explicit physical descriptor improves prediction of optical spectra, we augment the original GotenNet embedding with covalent radius gating.
Each atom's covalent radius $c_i \in \mathbb{R}$ is expanded into a fixed-width representation via a Gaussian RBF basis $\boldsymbol{\psi}(c_i)$, providing a continuous encoding of atomic size.

Neighbor messages are then formed by gating the projected species embedding with the neighbor's covalent radius embedding:
\begin{equation}
    \mathbf{m}_i=\sum_{j\in \mathcal{N}(i)}
    \Bigl(\sigma(\mathbf{W}_\text{nbr}\mathbf{z}_j)
    \circ \mathbf{W}_{\text{cbr}}\boldsymbol{\psi}(c_j)\Bigr)
    \circ \mathbf{d}_{ij},
\end{equation}
where
\begin{equation}
    \mathbf{d}_{ij}=
    f_{\text{cut}}(\|\mathbf{r}_{ij}\|)\cdot
    \mathbf{W}_{\text{ndp}}\,\boldsymbol{\varphi}(\|\mathbf{r}_{ij}\|)
\end{equation}
projects a exponential RBF expansion $\boldsymbol{\varphi}(\|\mathbf{r}_{ij}\|)$, cosine-cutoff weighted to suppress distant neighbors.
The source atom embedding is similarly gated by its own covalent radius:
\begin{equation}
    \mathbf{a}_i =
    \sigma(\mathbf{W}_{\text{na}}\mathbf{z}_i)
    \circ \mathbf{W}_{\text{cna}}\boldsymbol{\psi}(c_i).
\end{equation}
This multiplicative gating allows atomic size to modulate the contribution of species identity, encouraging the model to distinguish atoms that share a period or group but differ in effective radius.
Node features are then initialized by a two-layer MLP over the concatenation of $\mathbf{a}_i$ and $\mathbf{m}_i$:
\begin{equation}
    \mathbf{h}_{i,\text{init}}=
    \mathbf{W}_{\text{nru}}\,
    \sigma\!\left(\mathrm{LN}\!\left(
        \mathbf{W}_{\text{nrd}}[\mathbf{a}_i,\mathbf{m}_i]
    \right)\right)
    +\mathbf{b}_{\text{nru}},
\end{equation}
where $\sigma$ denotes SiLU and $\mathrm{LN}$ denotes layer normalization.
\paragraph{Scalar edge feature initialization.} The scalar edge features $\mathbf{t}_{ij} \in \mathbb{R}^{F}$ are initialized as the element-wise product of the summed initial node features and a linear projection of the RBF-expanded edge distances:
\begin{equation}
    \mathbf{t}_{ij,\text{init}}=(\mathbf{h}_{i,\text{init}}+\mathbf{h}_{j,\text{init}})\circ(\mathbf{W}_{\text{erp}}\varphi(||\mathbf{r}_{ij}||)+\mathbf{b}_{\text{erp}})
\end{equation}

\subsubsection{Message Block}
The message block updates node features via the Geometry-Aware Tensor Attention (GATA) module from GotenNet, adopted without modification except for removing the layer normalization on scalar node features. 

First, queries, keys, and a projected edge embedding are computed:
\begin{align}
    \mathbf{q}_i        & = \mathbf{W}_q\mathbf{h}_i + \mathbf{b}_q,                                           \\
    \mathbf{k}_j        & = \mathbf{W}_k\mathbf{h}_j + \mathbf{b}_k,                                           \\
    \mathbf{t}^{a}_{ij} & = \sigma\!\left(\mathbf{W}_\text{re}\,\mathbf{t}_{ij} + \mathbf{b}_\text{re}\right).
\end{align}
Splitting into $H$ heads of dimension $F_\text{head} = F/H$, the attention score for head $h$ is computed and normalized as:
\begin{align}
    \alpha_{ij}^{(h)} & = \bigl(\mathbf{q}_i^{(h)}\bigr)^\top\!\bigl(\mathbf{k}_j^{(h)} \circ \mathbf{t}_{ij}^{a,(h)}\bigr),                                                                   \\
    \alpha_{ij}^{(h)} & \leftarrow \frac{\exp\!\bigl(\alpha_{ij}^{(h)}\bigr)}{\sum_{k \in \mathcal{N}(i)} \exp\!\bigl(\alpha_{ik}^{(h)}\bigr)} \cdot \frac{\sqrt{|\mathcal{N}(i)|}}{\sqrt{F}},
\end{align}
where $|\mathcal{N}(i)|$ denotes the degree of node $i$. The value vector $\mathbf{v}_j = \gamma_v(\mathbf{h}_j) \in \mathbb{R}^{S \cdot F}$ is obtained via a two-layer MLP $\gamma_v \colon \mathbb{R}^F \to \mathbb{R}^{S \cdot F}$, with $S = 2L_\text{max} + 1$. Reshaping into $H$ heads of dimension $F_\text{ch} = S \cdot F / H$, the scaled edge attention (SEA) for head $h$ is:
\begin{equation}
    \mathbf{sea}_{ij}^{(h)} = \alpha_{ij}^{(h)}\,\mathbf{v}_j^{(h)} \in \mathbb{R}^{F_\text{ch}}.
\end{equation}

To compute the per-head coefficient vectors, we additionally introduce a second edge projection $\mathbf{t}^{c}_{ij} = \mathbf{W}_\text{rs}\,\mathbf{t}_{ij} + \mathbf{b}_\text{rs} \in \mathbb{R}^{S \cdot F}$ and an MLP-transformed neighbor node feature $\mathbf{h}^{c}_j = \gamma_s(\mathbf{h}_j) \in \mathbb{R}^{S \cdot F}$, both reshaped into $H$ heads so that $\mathbf{t}^{c,(h)}_{ij},\, \mathbf{h}^{c,(h)}_j \in \mathbb{R}^{F_\text{ch}}$:
\begin{equation}
    \mathbf{c}_{ij}^{(h)} = \mathbf{sea}_{ij}^{(h)} + f_\text{cut}\!\left(\|\mathbf{r}_{ij}\|\right) \cdot \bigl(\mathbf{t}^{c,(h)}_{ij} \circ \mathbf{h}^{c,(h)}_j\bigr) \in \mathbb{R}^{F_\text{ch}}.
\end{equation}
Concatenating across all heads and reshaping gives $\mathbf{c}_{ij} \in \mathbb{R}^{S \times F}$, which we decompose into $S$ components of dimension $F$:
\begin{equation}
    \mathbf{c}_{ij} = \Bigl[\mathbf{c}^s_{ij},\; \bigl\{\mathbf{c}^{d,(l)}_{ij}\bigr\}_{l=1}^{L_\text{max}},\; \bigl\{\mathbf{c}^{t,(l)}_{ij}\bigr\}_{l=1}^{L_\text{max}}\Bigr].
\end{equation}
These coefficients define the messages to the scalar and steerable node features:
\begin{align}
    \Delta \mathbf{h}_i                   & = \bigoplus_{j \in \mathcal{N}(i)} \mathbf{c}^s_{ij},                                                                                                                          \\
    \Delta \widetilde{\mathbf{X}}^{(l)}_i & = \bigoplus_{j \in \mathcal{N}(i)} \Bigl(\mathbf{c}^{d,(l)}_{ij} \circ \widetilde{\mathbf{r}}^{(l)}_{ij} + \mathbf{c}^{t,(l)}_{ij} \circ \widetilde{\mathbf{X}}^{(l)}_j\Bigr),
\end{align}
which are added to the respective features:
\begin{equation}
    \mathbf{h}_i \leftarrow \mathbf{h}_i + \Delta \mathbf{h}_i, \qquad \widetilde{\mathbf{X}}^{(l)}_i \leftarrow \widetilde{\mathbf{X}}^{(l)}_i + \Delta \widetilde{\mathbf{X}}^{(l)}_i.
\end{equation}

Note that the high-degree steerable features are initialized as \(\mathbf{X}^{(l)}\equiv \mathbf{0}\), which makes the \(\mathbf{c}^{t,(l)}_{ij}\circ \mathbf{\tilde{X}}^{(l)}_j\) term equal zero in the first message block, resulting in the initial features:

\begin{equation}
    \mathbf{\tilde{X}}^{(l)}_{i,\text{init}}=\bigoplus_{j\in \mathcal{N}(j)}(\mathbf{c}^{d,(l)}_{ij}\circ\mathbf{\tilde{r}}^{(l)}_{ij})
\end{equation}


\subsubsection{Update Block}
The update block updates the scalar edge features through the Hierarchical Tensor Refinement (HTR) module and the scalar and steerable node features through the Equivariant Feed-Forward (EQFF) network proposed in GotenNet.

\paragraph{Scalar edge feature update.} For each degree $l=1,\dots,L_{\text{max}}$, we compute transformed steerable features:
\begin{equation}
    \mathbf{Q}_i^{(l)}=\mathbf{\tilde{X}}_i^{(l)}\mathbf{W}_{vq}, \quad \mathbf{K}_j^{(l)}=\mathbf{\tilde{X}}_j^{(l)}\mathbf{W}_{vk}^{(l)},
\end{equation}
where $\mathbf{W}_{vq}$ is shared across degrees and $\mathbf{W}_{vk}^{(l)}$ is degree-specific. We then obtain $\hat{\mathbf{Q}}_i^{(l)}$ and $\hat{\mathbf{K}}_j^{(l)}$ by applying vector rejection of $\mathbf{Q}_i^{(l)}$ onto $\tilde{\mathbf{r}}_{ij}^{(l)}$ and of $\mathbf{K}_j^{(l)}$ onto $-\tilde{\mathbf{r}}_{ij}^{(l)}$, respectively. This ensures the outputs retain only what the network has learned beyond the edge geometry. 

Next an interaction vector $\mathbf{w}_{ij}$ is obtained by summing element-wise products across all degrees and components:
\begin{equation}
    \mathbf{w}_{ij} = \sum_{l=1}^{L_{\text{max}}} \sum_{m=1}^{2l+1} \hat{\mathbf{Q}}^{(l)}_{i,m} \circ \hat{\mathbf{K}}^{(l)}_{j,m},
\end{equation}
capturing learned atomic interactions not explained by edge geometry alone. Finally, the residual update to the scalar edge features is:
\begin{equation}
    \Delta \mathbf{t}_{ij}=\gamma_w(\mathbf{w}_{ij})\circ\gamma_t(\mathbf{t}_{ij}),
\end{equation}
where $\gamma_w$ and $\gamma_t$ are two-layer MLPs, giving $\mathbf{t}_{ij}\leftarrow \mathbf{t}_{ij} + \Delta \mathbf{t}_{ij}$.

\paragraph{Node feature updates. } For each \(l=1,\dots,L_{\text{max}}\) the steerable feature go through a linear projection \(\hat{\mathbf{X}_i}^{(l)}=\mathbf{\tilde{X}}_i^{(l)}\mathbf{W}_{\text{vu}}\).
Second, we compute a channel-wise norm over all projected features:

\begin{equation}
    \mathbf{n}_i=\sqrt{\sum_{l=1}^{L_{\text{max}}}\sum_{m=1}^{2l+1}(\hat{\mathbf{X}}_{i,m}^{(l)})^2+\epsilon}
\end{equation}

Next, the scalar node features and this norm are concatenated and passed through a two-layer MLP, which is split into two parts:

\begin{equation}
    [\mathbf{m}_{i,1},\mathbf{m}_{i,2}]=\gamma_m((\mathbf{n}_i,\mathbf{h}_i))
\end{equation}

After which, the scalar and high-degree steerable node features are updated as:
\begin{align}
    \mathbf{h}_i                 & \leftarrow \mathbf{m}_{1,i}+\mathbf{h}_i                                                      \\
    \mathbf{{\tilde{X}}}_i^{(l)} & \leftarrow (\mathbf{m}_{i,2} \circ \hat{\mathbf{X}}_{i}^{(l)})+\mathbf{{\tilde{X}}}_{i}^{(l)}
\end{align}

\subsubsection{Readout Layer}
For the readout layer, we use the same design as OptiMate3B. It consists of an 
attention-weighted global pooling to produce a graph-level embedding, followed by 
an output MLP to predict the spectrum. First, a linear layer with SiLU activation 
computes a per-node attention vector $\mathbf{a}_i = \sigma\!\left(\mathbf{W}_\text{pool}
\mathbf{h}_i + \mathbf{b}_\text{pool}\right) \in \mathbb{R}^{F}$ from the node features, 
which is then normalised across all nodes in the graph:
\begin{equation}
    \boldsymbol{\alpha}_i = \frac{\exp(\mathbf{a}_i)}{\sum_{j \in V} 
    \exp(\mathbf{a}_j)} \in \mathbb{R}^{F},
\end{equation}
where exp and division are applied elementwise, and the normalisation is taken 
over all nodes $j$ belonging to the same graph. The graph-level embedding is then 
the weighted sum:
\begin{equation}
    \mathbf{h}_G = \sum_{i \in V} \boldsymbol{\alpha}_i \circ \mathbf{h}_i 
    \in \mathbb{R}^{F}.
\end{equation}
Second, the graph embedding is passed through a final MLP, with $D$ hidden layers 
of dimension $F_s$, each followed by a SiLU activation (replacing the ReLU used in 
OptiMate3B), and a plain linear output layer. 

For the OptiMate3B dataset, the MLP produces 
$\hat{\mathbf{y}} = \gamma_\text{out}(\mathbf{h}_G) \in \mathbb{R}^{4002}$, where 
the first 2001 elements correspond to the imaginary part and the remaining elements 
to the real part. For the GNNOpt dataset, the output is instead 
$\hat{\mathbf{y}} = \gamma_\text{out}(\mathbf{h}_G) \in \mathbb{R}^{251}$, 
predicting either the real or imaginary part.


\section{Results}
\subsection{OptiMate3B Dataset}
\label{sec:experiments}
\begin{figure}[!htbp]
    \centering
    \includegraphics[width=\columnwidth]{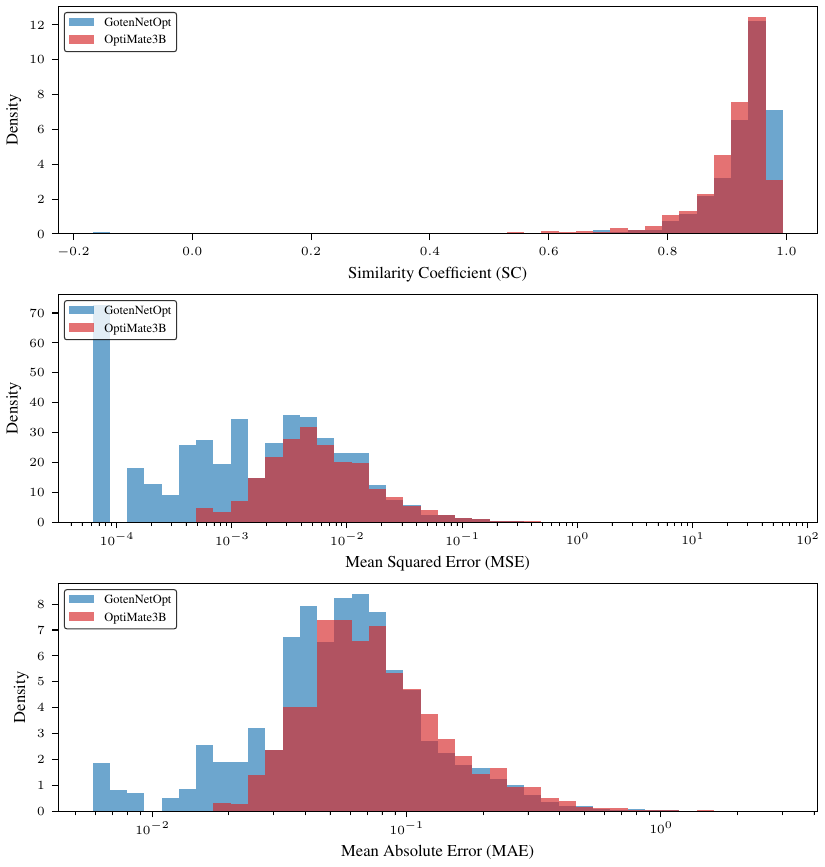}
    \caption{Density histograms of SC (top), MSE (middle), and MAE (bottom) across the RPA test samples for the two models. MSE and MAE are shown on logarithmic x-axes to accommodate their heavy-tailed distributions.}
    \label{fig:error_distributions}
\end{figure}

We follow the same procedure as OptiMate3B, training first on IPA spectra and finetuning on the RPA dataset. Training details and dataset descriptions are provided in \cref{app:train_Optimate3B,app:OptiMate3B_data}. Code to reproduce results is available at \href{https://github.com/khelverskovp/GotenNetOpt}{https://github.com/khelverskovp/GotenNetOpt}. We evaluate using three metrics over the frequency range $[0,\omega_{max}]$: the similarity coefficient (SC), mean squared error (MSE), and mean absolute error (MAE), defined as:

\begin{align}
    \text{SC}[X(\omega); Y(\omega)]  & = 1 - \frac{\int |X(\omega) - Y(\omega)|\, d\omega}{\int |Y(\omega)|\, d\omega}, \\[10pt]
    \text{MSE}[X(\omega); Y(\omega)] & = \frac{1}{\omega_{\max}} \int |X(\omega) - Y(\omega)|^2\, d\omega,              \label{eq:mse} \\[10pt]
    \text{MAE}[X(\omega); Y(\omega)] & = \frac{1}{\omega_{\max}} \int |X(\omega) - Y(\omega)|\, d\omega.
\end{align}

where $X(\omega)$ is the target spectra and $Y(\omega)$ is the predicted spectra for $\text{Im}(\bar{\varepsilon})$ or $\text{Re}(\bar{\varepsilon})$.
\begin{table*}[!t]
    \caption{IPA and RPA test set results for the imaginary (Im) and real (Re) parts of the dielectric function,
        measured by mean absolute error (MAE), mean squared error (MSE), and similarity coefficient (SC).
        Each cell shows the mean with median in parentheses below.
        Best results are shown in bold.}
    \label{results-table}
    \centering
    \resizebox{\textwidth}{!}{%
        \begin{sc}
            \begin{tabular}{p{2.5cm}cccccccccccc}
                \toprule
                      & \multicolumn{3}{c}{$\text{Im}(\bar{\varepsilon}_\text{IPA})$} & \multicolumn{3}{c}{$\text{Re}(\bar{\varepsilon}_\text{IPA})$} & \multicolumn{3}{c}{$\text{Im}(\bar{\varepsilon}_\text{RPA})$} & \multicolumn{3}{c}{$\text{Re}(\bar{\varepsilon}_\text{RPA})$}                                                                                                                                                         \\
                \cmidrule(r){2-4} \cmidrule(lr){5-7} \cmidrule(lr){8-10} \cmidrule(l){11-13}
                Model & MAE $\downarrow$ & MSE $\downarrow$ & SC $\uparrow$ & MAE $\downarrow$ & MSE $\downarrow$ & SC $\uparrow$ & MAE $\downarrow$ & MSE $\downarrow$ & SC $\uparrow$ & MAE $\downarrow$ & MSE $\downarrow$ & SC $\uparrow$ \\
                \midrule
                OptiMate3B
                      & 0.265                                                         & 0.744                                                         & 0.883                                                         & 0.276                                                         & 0.677            & 0.882            & 0.195            & 0.539            & 0.906            & 0.199            & 0.393            & 0.910            \\
                      & (0.193)                                                       & (0.108)                                                       & (0.903)                                                       & (0.201)                                                       & (0.108)          & (0.904)          & (0.115)          & (0.035)          & (0.930)          & (0.120)          & (0.035)          & (0.933)          \\
                \midrule
                $\text{GotenNet}_\text{Opt}$
                      & \textbf{0.238}                                                & \textbf{0.643}                                                & \textbf{0.894}                                                & \textbf{0.247}                                                & \textbf{0.601}   & \textbf{0.894}   & \textbf{0.164}   & \textbf{0.453}   & \textbf{0.919}   & \textbf{0.168}   & \textbf{0.318}   & \textbf{0.923}   \\
                      & \textbf{(0.173)}                                              & \textbf{(0.089)}                                              & \textbf{(0.913)}                                              & \textbf{(0.178)}                                              & \textbf{(0.089)} & \textbf{(0.914)} & \textbf{(0.093)} & \textbf{(0.024)} & \textbf{(0.942)} & \textbf{(0.099)} & \textbf{(0.024)} & \textbf{(0.946)} \\
                \bottomrule
            \end{tabular}
        \end{sc}
    }%
    \vskip -0.1in
\end{table*}
\newpage
\cref{results-table} reports mean and median values for each metric on both the IPA and RPA datasets, separated by imaginary and real parts. Additionally, we report results on using Voronoi graph construction and a scaled down version of \(\text{GotenNet}_\text{Opt}\) in \cref{app:results}.
Median values are considerably lower than means reflecting a tail of outlier samples. RPA predictions are substantially better than IPA for both models due to the fine-tuning procedure and the greater physical accuracy of RPA spectra. The distribution of each evaluation metric on the RPA test set are shown in \cref{fig:error_distributions}.

The difference in distributions is most pronounced for MAE and MSE, particularly the latter, as evidenced by the performance summary in ~\cref{tab:improvement}. We report percentage improvements for both the full spectrum (0–20 eV) and the low-energy sub-spectrum (0–8 eV), which is most relevant for thin-film optics. The relative performance improvements are largest on the RPA dataset and in the 0–8 eV sub-spectrum. 

\begin{table}[htbp]
\caption{Percentage improvement of $\text{GotenNet}_\text{Opt}$ over OptiMate3B in median MAE, MSE, and SC across the test set, reported for the full $0$--$20$\,eV range and the low-energy $0$--$8$\,eV sub-range.}
\label{tab:improvement}
\centering
\resizebox{\columnwidth}{!}{%
\begin{sc}
\begin{tabular}{llcc}
\toprule
 & Metric & ${\Delta\%}$ \text{(0--20 \textup{eV})} & ${\Delta\%}$ \text{(0--8 \textup{eV})} \\
\midrule
\multirow{3}{*}{$\text{Im}(\bar{\varepsilon}_\text{IPA})$}
    & MAE $\downarrow$ & $10.1\%$ & $8.9\%$  \\
    & MSE $\downarrow$ & $18.0\%$ & $16.0\%$ \\
    & SC $\uparrow$  & $1.1\%$  & $1.6\%$  \\
\midrule
\multirow{3}{*}{$\text{Re}(\bar{\varepsilon}_\text{IPA})$}
    & MAE $\downarrow$ & $11.0\%$ & $11.4\%$ \\
    & MSE $\downarrow$ & $17.9\%$ & $18.2\%$ \\
    & SC  $\uparrow$  & $1.1\%$  & $1.1\%$  \\
\midrule
\multirow{3}{*}{$\text{Im}(\bar{\varepsilon}_\text{RPA})$}
    & MAE $\downarrow$ & $18.4\%$ & $19.2\%$ \\
    & MSE $\downarrow$ & $31.2\%$ & $31.3\%$ \\
    & SC $\uparrow$ & $1.2\%$  & $1.4\%$  \\
\midrule
\multirow{3}{*}{$\text{Re}(\bar{\varepsilon}_\text{RPA})$}
    & MAE $\downarrow$ & $17.9\%$ & $21.5\%$ \\
    & MSE $\downarrow$ & $31.6\%$ & $33.2\%$ \\
    & SC  $\uparrow$  & $1.4\%$  & $1.3\%$  \\
\midrule
\multirow{2}{*}{Average}
    & MAE/MSE $\downarrow$ & $\mathbf{19.6\%}$ & $\mathbf{20.0\%}$ \\
    & SC $\uparrow$        & $\mathbf{1.2\%}$  & $\mathbf{1.4\%}$  \\
\bottomrule
\end{tabular}%
\end{sc}
}
\end{table}
\newpage
Table~\ref{tab:static_permittivity} compares both models on predicting the static real permittivity by evaluating the median MAE and MSE on $\text{Re}(\bar{\varepsilon}(0))$ from the predicted spectra for both the IPA and RPA test sets.

\begin{table}[!htbp]
\caption{Test set MAE and MSE (median) on the static real permittivity, and the percentage improvement of \(\text{GotenNet}_\text{Opt}\).}
\label{tab:static_permittivity}
\centering
\resizebox{\columnwidth}{!}{%
\begin{sc}
\begin{tabular}{lcccc}
\toprule
\multirow{2}{*}{Model}
    & \multicolumn{2}{c}{$\text{Re}(\bar{\varepsilon}(0))_\mathrm{IPA}$}
    & \multicolumn{2}{c}{$\text{Re}(\bar{\varepsilon}(0))_\mathrm{RPA}$} \\
\cmidrule(lr){2-3} \cmidrule(lr){4-5}
    & MAE $\downarrow$ & MSE $\downarrow$
    & MAE $\downarrow$ & MSE $\downarrow$ \\
\midrule
OptiMate3B   & 0.131   & 0.017   & 0.121   & 0.015  \\
$\text{GotenNet}_\text{Opt}$  & 0.080   & 0.006   & 0.074   & 0.006  \\
\midrule
\textbf{$\Delta\%$}
                      & $\mathbf{39.0\%}$ & $\mathbf{62.8\%}$ & $\mathbf{38.5\%}$ & $\mathbf{62.1\%}$ \\
\bottomrule
\end{tabular}%
\end{sc}
}
\end{table}

\begin{figure*}[!htbp]
    \centering
    \includegraphics[width=0.90\textwidth]{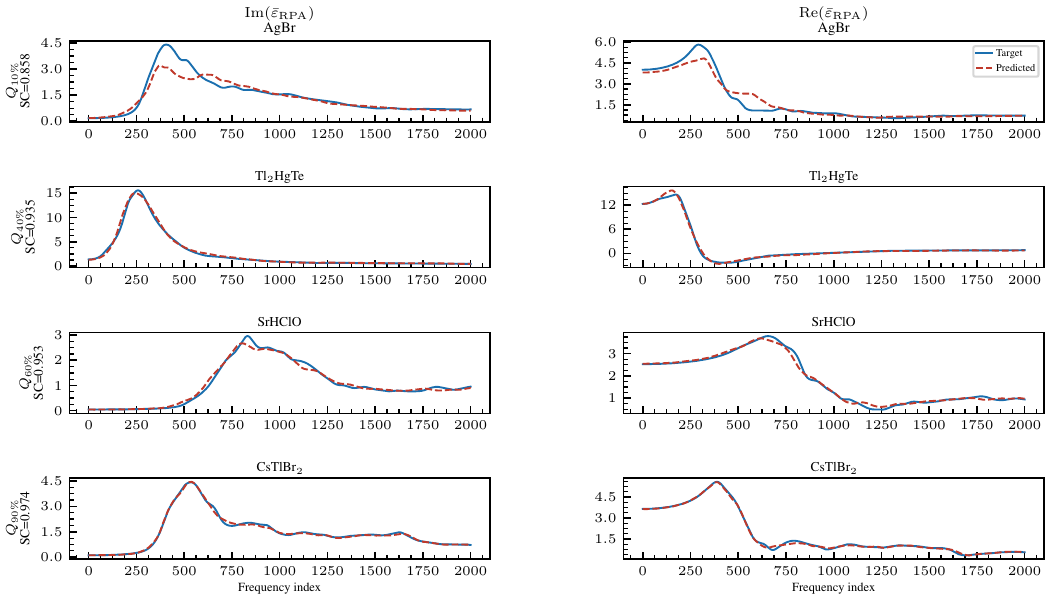}
    \caption{Predicted (dashed, red) vs.\ target (solid, blue) dielectric
        spectra at the 10th, 40th, 60th, and 90th percentiles of the RPA test set
        ranked by SC. Each row shows
        $\mathrm{Im}(\bar{\varepsilon}_{\mathrm{RPA}})$ and
        $\mathrm{Re}(\bar{\varepsilon}_{\mathrm{RPA}})$ components for a single
        structure (chemical formula shown above).}
    \label{fig:quantile-spectra}
\end{figure*}

Figure~\ref{fig:quantile-spectra} shows predicted versus target spectra at the 10th, 40th, 60th and 90th SC percentiles of the RPA test set. For additional spectra plots, Figures~\ref{fig:top5_gotenNetWins}--\ref{fig:top5_bothGood} in \cref{app:add_plots} show representative spectra for
the cases where \(\text{GotenNet}_\text{Opt}\) outperforms OptiMate3B, where OptiMate3B
outperforms \(\text{GotenNet}_\text{Opt}\), and where both models perform worst and best simultaneously. 
The imaginary and real parts are predicted with comparable accuracy across percentiles, consistent with the Kramers-Kronig relations.

\begin{figure}[htbp]
    \centering
    \includegraphics[width=0.8\columnwidth]{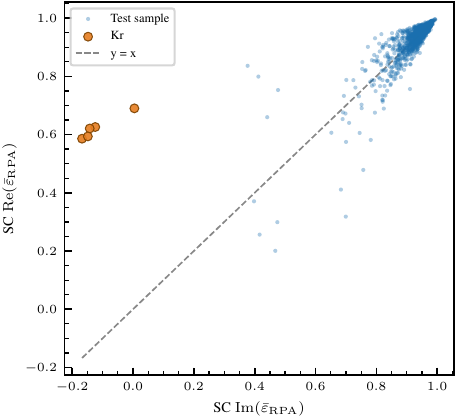}
    \caption{Kramers-Kronig consistency check. Per-sample SC for \(\text{Re}(\bar{\varepsilon})\) against SC for \(\text{Im}(\bar{\varepsilon})\).}
    \label{fig:kkplot}
\end{figure}

\newpage
Figure~\ref{fig:kkplot} confirms this explicitly, having a strong-per sample correlation between SC on $\text{Re}(\bar{\varepsilon})$ and $\text{Im}(\bar{\varepsilon})$. 

\begin{figure*}[htbp]
    \centering
    \includegraphics[width=\textwidth]{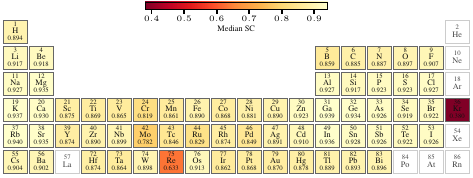}
    \vspace{0.1in}
    \caption{Median SC for each compound in the IPA test set that contains a given element. Elements that are not present in any compounds are colored white.}
    \label{fig:periodic_table}
\end{figure*}

\begin{figure}[htbp]
    \centering
    \includegraphics[width=0.9\columnwidth]{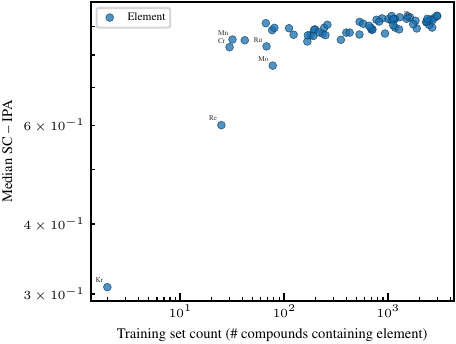}
    \caption{Log-log plot of median SC for all compounds in the test set that contain a given element as a function of how often that element occurs in the training set for the IPA data.}
    \label{fig:sc_v_count}
\end{figure}

Five outliers are Kr-only crystals. Kr appears in only 2 training samples with no other group-18 elements (Figure~\ref{fig:periodic_table}), leaving its atom embedding poorly informed; Kr also forms no covalent bonds under ordinary conditions, meaning the covalent radius feature is physically meaningless. These outliers are an extreme case of a broader log-linear relationship between prediction accuracy and per-element training count (\cref{fig:sc_v_count}), consistent with findings reported for OptiMate3B~\citep{OptiMate3B}.


\subsection{GNNOpt Dataset}
\cref{tab:comparisonGNNOpt} presents the performance of $\text{GotenNet}_\text{Opt}$ on the GNNOpt dataset; training details and dataset descriptions are provided in~\cref{app:train_gnnopt,app:GNNOpt_data}. Following~\citet{GNNOpt}, we report MSE and $R^2$ on the test set. $R^2$ is evaluated on the 
energy-weighted mean $\bar{X}_i = \sum_k \omega_k X_i(\omega_k) / \sum_k \omega_k$ of each 
predicted and target spectrum,
\begin{equation}
    R^2[X; Y] = 1 - \frac{\sum_{i=1}^{N}\left(\bar{X}_i - \bar{Y}_i\right)^2}{\sum_{i=1}^{N}\left(\bar{X}_i - \overline{X}\right)^2},
\end{equation}
where $\bar{Y}_i$ is defined analogously and $\overline{X}$ is the mean of $\bar{X}_i$ over the 
test set. MSE is computed over all $K=251$ frequency points, consistent with the 
definition in \cref{eq:mse}.
\begin{table}[htbp]
\caption{Test set results on the GNNOpt dataset measured by MSE and coefficient of determination (\(R^2\)) along with percentage improvement compared to GNNOpt.}
\label{tab:comparisonGNNOpt}
\centering
\resizebox{\columnwidth}{!}{%
\begin{sc}
\begin{tabular}{llccc}
\toprule
 & Metric & GNNOpt & $\text{GotenNet}_\text{Opt}$ & $\Delta\%$ \\
\midrule
\multirow{2}{*}{$\text{Im}(\bar{\varepsilon})$}
    & MSE $\downarrow$ & $1.765$& $1.371$& $22.3\%$\\
    & $R^2$ $\uparrow$ & $0.737$ & $0.864$ & $17.2\%$  \\
\midrule
\multirow{2}{*}{$\text{Re}(\bar{\varepsilon})$}
    & MSE $\downarrow$ & $2.363$ & $1.907$ & $19.3\%$ \\
    & $R^2$ $\uparrow$ & $0.715$ & $0.806$ & $12.7\%$ \\
\bottomrule
\end{tabular}%
\end{sc}
}
\end{table}

\section{Discussion}
This work suggests that equivariant GNNs can meaningfully improve optical spectra prediction, with the most pronounced gains in the 0--8 eV range and for prediction of the static real permittivity. Gains are also proportionally larger on the RPA dataset than on the IPA dataset, which may reflect that physically accurate spectra encoding richer many-body interactions benefit more from expressive geometric representations

The adaptation of GotenNet involved two main changes: incorporating covalent radius embeddings into the scalar node feature initialization, and replacing the readout layer with the OptiMate3B readout design. Covalent radius is a physically motivated feature, though it is not universally informative, and more generally, the most useful atomic features may depend on the chemical subspace being screened. Future work could explore other physical properties as node features, tailored to the target application. The readout architecture is similarly not necessarily optimal, and other designs remain to be explored. More broadly, other equivariant GNN architectures have not yet been tested on this task. 

Despite performance improvements, the dominant failure mode is still distributional: prediction accuracy correlates with per-element training frequency. Discarding these outliers, the similarity coefficient metric suggests the model is approaching highly accurate predictions on the current data. Hence, for elements well represented in the training data, the model is viable as a large-scale screening tool, but for structures containing underrepresented elements, predictions should be treated with caution, and expanding the training set would be necessary before using the model on poorly covered chemical spaces. 

\section{Conclusion}
We have presented $\text{GotenNet}_\text{Opt}$, an adaptation of the equivariant GNN GotenNet for optical spectra prediction. The results show that using an equivariant architecture leads to consistent performance increases, indicating that geometric equivariance is a beneficial inductive bias for this task. Gains are most pronounced in the low-energy range and in the static real permittivity, quantities of direct relevance for high-throughput screening of thin-film optical materials, and are proportionally larger on the RPA dataset, suggesting that geometric expressiveness matters most when the training data is physically accurate.


\section*{Impact Statement}
This paper presents work whose goal is to advance the field of Machine
Learning. There are many potential societal consequences of our work, none
which we feel must be specifically highlighted here.

\bibliography{example_paper}
\bibliographystyle{icml2026}

\newpage
\appendix
\onecolumn
\section{Density functional theory}
\label{app:dft_background}
Optical response tensors of materials are calculated through quantum mechanical perturbation theory which necessitates an accurate description of the electronic ground state of the material to be perturbed first. For many-body systems, that is, systems containing many electrons, the ground state of the system is typically obtained through DFT. Strictly speaking, determining the ground state of a quantum system with $N$ electrons means determining the ground state many-body wave function $\Psi(\mathbf{r}_1,\mathbf{r}_2,\ldots,\mathbf{r}_N)$; however, within DFT the ground state can be understood entirely through the ground state electron density $\rho(\mathbf{r})$ which is otherwise considered an observable that can be calculated directly from the many-body wave function. While many-body wave functions describe the ground states of interacting electrons governed by the Schrödinger equation, most modern formulations of DFT use an auxiliary set of single-particle wave functions which obey the Kohn-Sham equations~\cite{kohn-sham}. These single-particle wave functions describe a fictitious set of non-interacting electrons; however, the Kohn-Sham equations are cleverly constructed such that they (in theory) yield the same electron density as the true many-body wave function. Density functional theory thus discards the staggeringly complicated many-body wave function in favor of the much simpler set of single-particle wave functions as both will construct the true ground state electron particle density $\rho$.

Although DFT is a popular and versatile method for obtaining ground state information, trouble arises when considering excited states of the system as only the ground state can be described entirely by the density, while excited states need the true many-body wave function. This leads to a set of perturbation theory methods tiered by how diligently they account for many-body effects, which are otherwise missing in the non-interacting Kohn-Sham system. The simplest of which is the independent-particle approximation, which simply treats the optical response as a sum of optical responses of the independent-particle system.

\section{Training Details}
\label{app:training_details}
\subsection{Training on the OptiMate3B Dataset}
\label{app:train_Optimate3B}
Table~\ref{tab:hyperparams} shows the hyperparameters used for training on the OptiMate3B dataset. Training proceeds in two stages. In the first stage, the model is trained to predict IPA
dielectric spectra. In the second stage, the best IPA checkpoint is used to initialise
the model, which is then fine-tuned on RPA dielectric spectra.

Both stages use the AdamW optimiser with a learning rate of $1 \times 10^{-5}$ and weight
decay of $0.01$. Gradients are clipped to a maximum norm of $10$. The learning rate is
reduced on plateau by a factor of $0.5$, with a minimum learning rate of $10^{-8}$ and
$10^{-9}$ for IPA and RPA training respectively. The plateau scheduler patience is set to
$15$ epochs. Early
stopping is applied with a patience of $50$ epochs in both stages.

The loss function is mean absolute error (MAE) in both stages. To stabilize early stopping,
an exponential moving average (EMA) of the validation loss is maintained with smoothing
factor $\alpha = 0.9$. Validation loss spikes exceeding $2\times$ the current smoothed
value are ignored when updating the EMA, preventing transient instabilities from triggering
premature learning rate decay or stopping.
\begin{table}[htbp]
    \centering
    \caption{Hyperparameters for the two-stage training procedure (IPA pre-training and RPA fine-tuning). Shared architecture parameters apply to both stages. (S) are parameter values used for the smaller variant.}
    \label{tab:hyperparams}
    \begin{tabular}{lll}
        \toprule
        \textbf{Category} & \textbf{Hyperparameter}      & \textbf{Value}                                 \\
        \midrule
        \multirow{6}{*}{Optimisation}
                          & Learning rate                & $1 \times 10^{-5}$ (S: $1 \times 10^{-4}$) \\
                          & Minimum learning rate        & $1 \times 10^{-8}$ (RPA: $1 \times 10^{-9}$)   \\
                          & Weight decay                 & $0.01$                                         \\
                          & Gradient clipping            & $10$                                           \\
                          & Batch size                   & $32$                                           \\
                          & Max epochs                   & $1000$                                         \\
        \midrule
        \multirow{4}{*}{Scheduler \& Stopping}
                          & LR scheduler                 & ReduceLROnPlateau                              \\
                          & LR reduction factor          & $0.5$                                          \\
                          & LR patience                  & $15$ epochs                   \\
                          & Early stopping patience      & $50$ epochs                                   \\
        \midrule
        \multirow{3}{*}{Loss}
                          & Loss function                & MAE                                            \\
                          & Val.\ loss EMA $\alpha$      & $0.9$                                          \\
                          & Spike filter factor          & $2.0$                                          \\
        \midrule
        \multirow{3}{*}{Data Split}
                          & Training fraction            & $80\%$                                         \\
                          & Validation fraction          & $10\%$                                         \\
                          & Test fraction                & $10\%$                                         \\
        \midrule
        \multirow{9}{*}{Architecture}
                          & Feature embedding dim ($F$)  & $1024$ (S: $256$)                          \\
                          & Message passing rounds ($T$) & $6$                                            \\
                          & Radial basis functions       & $64$                                           \\
                          & Cutoff radius                & $6.0$ \AA                                      \\
                          & Max degree $(L_{\max})$        & $2$                                            \\
                          & Attention heads ($H$)        & $8$                                            \\
                          & Dropout                      & $0.1$                                          \\
                          & Readout depth ($D$)          & $4$ (S: $2$)                               \\
                          & Readout hidden dim ($F_s$)   & $8192$ (S: $256$)                          \\
        \bottomrule
    \end{tabular}
\end{table}

\subsection{Training on the GNNOpt dataset}
\label{app:train_gnnopt}
\cref{tab:hyperparams_gnnopt} shows the hyperparameters used for training on the GNNOpt dataset. We follow the GNNOpt training setup, including their graph construction procedure and the scaling and interpolation of the target spectra. Unlike the OptiMate3B dataset setup, where the real and imaginary parts of the dielectric function are predicted jointly, here a separate model is trained and evaluated for each component. The optical spectra are interpolated onto a uniform grid of 251 points over the energy range $0 \leq \hbar\omega \leq 50$\,eV. Spectra are then normalized by a global scale factor computed as the median of the per-sample maximum values across the training set. The loss function is mean squared error (MSE) on the normalized spectra.

Graphs are constructed by placing edges between all atoms within a cutoff radius of $r_{c} = 6.0$\,\AA. All experiments use a fixed random seed of 12 (same seed used in the original GNNOpt experiments). Training is performed for 200 epochs with a batch size of 8, and gradients are clipped to a maximum $\ell_2$ norm of 1.0.

\begin{table}[htbp]
    \centering
    \caption{Hyperparameters for training on the GNNOpt dataset.}
    \label{tab:hyperparams_gnnopt}
    \begin{tabular}{lll}
        \toprule
        \textbf{Category} & \textbf{Hyperparameter}                      & \textbf{Value}     \\
        \midrule
        \multirow{6}{*}{Optimisation}
                          & Optimizer                                    & AdamW              \\
                          & Learning rate                                & $6\times 10^{-4}$  \\
                          & Weight decay                                 & $0.015$            \\
                          & Gradient clipping                            & $1.0$              \\
                          & Batch size                                   & $8$                \\
                          & Epochs                                       & $200$              \\
        \midrule
        \multirow{2}{*}{Scheduler}
                          & LR scheduler                                 & Exponential        \\
                          & LR decay ($\gamma$)                          & $0.98$             \\
        \midrule
        \multirow{2}{*}{Loss \& Seed}
                          & Loss function                                & MSE                \\
                          & Random seed                                  & $12$               \\
        \midrule
        \multirow{9}{*}{Architecture}
                          & Feature embedding dim\ $(F)$     & $256$              \\
                          & Message passing rounds \((T)\)                             & $3$                \\
                          & Radial basis functions                       & $64$               \\
                          & Cutoff radius                   & $6.0$\,\AA         \\
                          & Max degree ($L_{\max}$) & $2$                \\
                          & Attention heads \((H)\)                              & $8$                \\
                          & Dropout                                      & $0.3$              \\
                          & Readout depth \((D)\)                                        & $1$                \\
                          & Readout hidden dim \((F_s)\)                                  & $256$              \\
        \bottomrule
    \end{tabular}
\end{table}

\section{Dataset Descriptions}
\label{app:dataset_description}
\subsection{OptiMate3B Dataset}
\label{app:OptiMate3B_data}
The OptiMate3B datasets consists of an IPA dataset with 23,834 materials and an RPA dataset with 10,533 materials, which we  obtain from \citet{Grunert2026Data1} and \citet{Grunert2026Data2}. The structures were originally drawn from the Alexandria database and filtered according to the following criteria:
\begin{itemize}
    \item no elements beyond Bismuth and no Lanthanides;
    \item a direct or indirect bandgap above 50 meV;
    \item an indirect bandgap below 100 meV or above 10 eV
    \item energy above hull below 150 meV per atom;
    \item symmetry of at least space group 75 (tetragonal or higher);
    \item at most 8 atoms per primitive unit cell.
\end{itemize}

For the RPA dataset, structures were further restricted to at most 4 atoms per unit cell. We adopt the same split procedure as \citet{OptiMate3B}, splitting by unique chemical composition in an 80:10:10 ratio yielding the exact same splits as in the original paper of 19,036, 2413, and 2385 materials for training, validation and test sets respectively. The
same compositional split is preserved for the RPA dataset, giving 8,414, 1,040, and 1,079 materials. 
\newpage
\subsection{GNNOpt Dataset}
\label{app:GNNOpt_data}
The GNNOpt dataset consists of 944 crystal structures obtained from the \href{https://github.com/nguyen-group/GNNOpt}{GNNOpt GitHub repository}. The optical spectra were originally computed using IPA by~\cite{yang2022highthroughputopticalabsorptionspectra}, with structures sourced from the Materials Project~\cite{MP}. The dataset was constructed by filtering for materials satisfying the following criteria:
\begin{itemize}
\item an energy band gap between 0 and 5.0 eV;
\item at most 10 atoms per primitive unit cell.
\end{itemize}
We adopt the same train/validation/test split as the original GNNOpt paper, taken directly from the repository, corresponding to an 80\%, 10\%, 10\% division of 733, 97, and 110 materials, respectively.

\section{Additional Results and Analysis}
\label{app:results}
\subsection{Covalent Radius Embeddings Ablation}
The effect of covalent radius embeddings is modest but generally positive, as shown in
Table~\ref{tab:gotennetopt_covalent}. The clearest gains are in the RPA targets, where MAE and
MSE improve consistently (up to $3.13\%$ and $4.00\%$ respectively). For the IPA targets the
picture is more mixed; MAE improves slightly but the median MSE is marginally worse ($+1.14\%$),
suggesting that covalent radius information is more beneficial for the RPA computed dielectric function.
Overall, the average median improvement across MAE and MSE is $1.18\%$ and $0.14\%$ for SC.

\begin{table*}[htbp]
    \caption{IPA and RPA test set results ablating covalent radius embeddings in $\text{GotenNet}_\text{Opt}$. Each cell shows the mean
    with median in parentheses below. Best results are shown in bold. The $\Delta\%$ row shows the
    relative change of $\text{GotenNet}_\text{Opt}$ compared to the model without covalent radius
    embeddings based on the medians.}
    \label{tab:gotennetopt_covalent}
    \centering
    \resizebox{\textwidth}{!}{%
        \begin{sc}
            \begin{tabular}{p{3cm}cccccccccccc}
                \toprule
                      & \multicolumn{3}{c}{$\text{Im}(\bar{\varepsilon}_\text{IPA})$} & \multicolumn{3}{c}{$\text{Re}(\bar{\varepsilon}_\text{IPA})$} & \multicolumn{3}{c}{$\text{Im}(\bar{\varepsilon}_\text{RPA})$} & \multicolumn{3}{c}{$\text{Re}(\bar{\varepsilon}_\text{RPA})$} \\
                \cmidrule(r){2-4} \cmidrule(lr){5-7} \cmidrule(lr){8-10} \cmidrule(l){11-13}
                Model & MAE $\downarrow$ & MSE $\downarrow$ & SC $\uparrow$ & MAE $\downarrow$ & MSE $\downarrow$ & SC $\uparrow$ & MAE $\downarrow$ & MSE $\downarrow$ & SC $\uparrow$ & MAE $\downarrow$ & MSE $\downarrow$ & SC $\uparrow$ \\
                \midrule
                w/o cov. radius
                      & 0.241          & 0.664          & 0.893          & 0.251          & 0.623          & 0.893          & 0.167          & 0.465          & 0.920          & 0.171          & 0.332          & 0.923          \\
                      & (0.174)        & (0.088)        & (0.912)        & (0.180)        & (0.088)        & (0.913)        & (0.096)        & (0.025)        & (0.940)        & (0.102)        & (0.024)        & (0.945)        \\
                \midrule
                $\text{GotenNet}_\text{Opt}$
                      & \textbf{0.238} & \textbf{0.643} & \textbf{0.894} & \textbf{0.247} & \textbf{0.601} & \textbf{0.894} & \textbf{0.164} & \textbf{0.453} & \textbf{0.919} & \textbf{0.168} & \textbf{0.318} & \textbf{0.923} \\
                      & \textbf{(0.173)} & \textbf{(0.089)} & \textbf{(0.913)} & \textbf{(0.178)} & \textbf{(0.089)} & \textbf{(0.914)} & \textbf{(0.093)} & \textbf{(0.024)} & \textbf{(0.942)} & \textbf{(0.099)} & \textbf{(0.024)} & \textbf{(0.946)} \\
                \midrule
                $\Delta\%$
                      & $0.57\%$ & $+1.14\%$ & $0.11\%$ & $1.11\%$ & $+1.14\%$ & $0.11\%$ & $3.13\%$ & $4.00\%$ & $0.21\%$ & $2.94\%$ & $0.00\%$ & $0.11\%$ \\
                \bottomrule
            \end{tabular}
        \end{sc}
    }%
    \vskip -0.1in
\end{table*}

\subsection{Effect of Model Size on Performance}
\label{sec:model_size}
Table~\ref{tab:gotennetopt_s_full} reports performance on the OptiMate3B dataset of $\text{GotenNet}_\text{Opt}\text{-S}$, a 
smaller variant (16M parameters) using readout depth 2, feature dimension 256, and readout hidden 
dimension 256, compared to $\text{GotenNet}_\text{Opt}$ (474M).
Despite its reduced size, $\text{GotenNet}_\text{Opt}\text{-S}$ trails $\text{GotenNet}_\text{Opt}$ 
by only 6.9\% in median MAE/MSE and 0.3\% in median SC on average, suggesting that performance gains stem primarily from 
architectural differences rather than increased parameter count. Nevertheless, since inference is run once per 
crystal in the intended screening use case, the computational overhead of the larger model
is negligible, making it the preferred choice over the smaller variant.
\begin{table*}[htbp]
    \caption{IPA and RPA test set results comparing $\text{GotenNet}_\text{Opt}\text{-S}$ (small variant)
    and $\text{GotenNet}_\text{Opt}$ (main model). Each cell shows the mean with median in parentheses below.
    Best results are shown in bold. The $\Delta\%$ row shows the relative improvement of $\text{GotenNet}_\text{Opt}$ compared to $\text{GotenNet}_\text{Opt}\text{-S}$ based on the medians.}
    \label{tab:gotennetopt_s_full}
    \centering
    \resizebox{\textwidth}{!}{%
        \begin{sc}
            \begin{tabular}{p{2.5cm}cccccccccccc}
                \toprule
                      & \multicolumn{3}{c}{$\text{Im}(\bar{\varepsilon}_\text{IPA})$} & \multicolumn{3}{c}{$\text{Re}(\bar{\varepsilon}_\text{IPA})$} & \multicolumn{3}{c}{$\text{Im}(\bar{\varepsilon}_\text{RPA})$} & \multicolumn{3}{c}{$\text{Re}(\bar{\varepsilon}_\text{RPA})$} \\
                \cmidrule(r){2-4} \cmidrule(lr){5-7} \cmidrule(lr){8-10} \cmidrule(l){11-13}
                Model & MAE $\downarrow$ & MSE $\downarrow$ & SC $\uparrow$ & MAE $\downarrow$ & MSE $\downarrow$ & SC $\uparrow$ & MAE $\downarrow$ & MSE $\downarrow$ & SC $\uparrow$ & MAE $\downarrow$ & MSE $\downarrow$ & SC $\uparrow$ \\
                \midrule
                $\text{GotenNet}_\text{Opt}\text{-S}$
                      & 0.243          & 0.679          & 0.893          & 0.252          & 0.640          & 0.893          & 0.170          & 0.452          & 0.918          & 0.173          & 0.326          & 0.921          \\
                      & (0.178)        & (0.092)        & (0.910)        & (0.183)        & (0.092)        & (0.913)        & (0.102)        & (0.028)        & (0.938)        & (0.105)        & (0.028)        & (0.943)        \\
                \midrule
                $\text{GotenNet}_\text{Opt}$
                      & \textbf{0.238} & \textbf{0.643} & \textbf{0.894} & \textbf{0.247} & \textbf{0.601} & \textbf{0.894} & \textbf{0.164} & \textbf{0.453} & \textbf{0.919} & \textbf{0.168} & \textbf{0.318} & \textbf{0.923} \\
                      & \textbf{(0.173)} & \textbf{(0.089)} & \textbf{(0.913)} & \textbf{(0.178)} & \textbf{(0.089)} & \textbf{(0.914)} & \textbf{(0.093)} & \textbf{(0.024)} & \textbf{(0.942)} & \textbf{(0.099)} & \textbf{(0.024)} & \textbf{(0.946)} \\
                \midrule
                $\Delta\%$
                      & $2.81\%$ & $3.26\%$ & $0.33\%$ & $2.73\%$ & $3.26\%$ & $0.11\%$ & $8.82\%$ & $14.29\%$ & $0.43\%$ & $5.71\%$ & $14.29\%$ & $0.32\%$ \\
                \bottomrule
            \end{tabular}
        \end{sc}
    }%
    \vskip -0.1in
\end{table*}

\subsection{Effect of Graph Construction on Performance}
We compare \(\text{GotenNet}_\text{Opt}\) trained with Voronoi-based graph construction against the
distance-based cutoff of 6 Å on the OptiMate3B dataset. As shown in Table~\ref{tab:gotennetopt_voronoi}, the cutoff-based
method consistently outperforms the Voronoi-based method across all targets and metrics, with an
average combined median improvement of $4.3\%$ in MAE and MSE and $0.3\%$ in SC. While the improvements
are modest, they are consistent across all four targets and both the IPA and RPA datasets,
suggesting that the fixed-radius cutoff provides a slightly better graph structure for this task
than the Voronoi approach.

\begin{table*}[htbp]
    \caption{IPA and RPA test set results comparing \(\text{GotenNet}_\text{Opt}\) on Voronoi vs. distance based graph construction. Each cell shows the mean with median in parentheses below.
    Best results are shown in bold. The $\Delta\%$ row shows the relative change of the 6~\AA\ cutoff method compared to Voronoi based on the medians.}
    \label{tab:gotennetopt_voronoi}
    \centering
    \resizebox{\textwidth}{!}{%
        \begin{sc}
            \begin{tabular}{p{2.5cm}cccccccccccc}
                \toprule
                      & \multicolumn{3}{c}{$\text{Im}(\bar{\varepsilon}_\text{IPA})$} & \multicolumn{3}{c}{$\text{Re}(\bar{\varepsilon}_\text{IPA})$} & \multicolumn{3}{c}{$\text{Im}(\bar{\varepsilon}_\text{RPA})$} & \multicolumn{3}{c}{$\text{Re}(\bar{\varepsilon}_\text{RPA})$} \\
                \cmidrule(r){2-4} \cmidrule(lr){5-7} \cmidrule(lr){8-10} \cmidrule(l){11-13}
                Model & MAE $\downarrow$ & MSE $\downarrow$ & SC $\uparrow$ & MAE $\downarrow$ & MSE $\downarrow$ & SC $\uparrow$ & MAE $\downarrow$ & MSE $\downarrow$ & SC $\uparrow$ & MAE $\downarrow$ & MSE $\downarrow$ & SC $\uparrow$ \\
                \midrule
                Voronoi
                      & 0.243          & 0.678          & 0.892          & 0.254          & 0.635          & 0.892          & 0.169          & 0.497          & 0.917          & 0.173          & 0.352          & 0.921          \\
                      & (0.178)        & (0.094)        & (0.910)        & (0.185)        & (0.094)        & (0.911)        & (0.099)        & (0.026)        & (0.939)        & (0.103)        & (0.026)        & (0.944)        \\
                \midrule
                6 Å cutoff
                      & \textbf{0.238} & \textbf{0.643} & \textbf{0.894} & \textbf{0.247} & \textbf{0.601} & \textbf{0.894} & \textbf{0.164} & \textbf{0.453} & \textbf{0.919} & \textbf{0.168} & \textbf{0.318} & \textbf{0.923} \\
                      & \textbf{(0.173)} & \textbf{(0.089)} & \textbf{(0.913)} & \textbf{(0.178)} & \textbf{(0.089)} & \textbf{(0.914)} & \textbf{(0.093)} & \textbf{(0.024)} & \textbf{(0.942)} & \textbf{(0.099)} & \textbf{(0.024)} & \textbf{(0.946)} \\
                \midrule
                $\Delta\%$
                      & $2.81\%$ & $5.32\%$ & $0.33\%$ & $3.78\%$ & $5.32\%$ & $0.33\%$ & $6.06\%$ & $7.69\%$ & $0.32\%$ & $3.88\%$ & $7.69\%$ & $0.21\%$ \\
                \bottomrule
            \end{tabular}
        \end{sc}
    }%
    \vskip -0.1in
\end{table*}

\subsection{Additional spectra plots}
\label{app:add_plots}
\begin{figure*}[t]
    \centering
    \includegraphics[width=\textwidth]{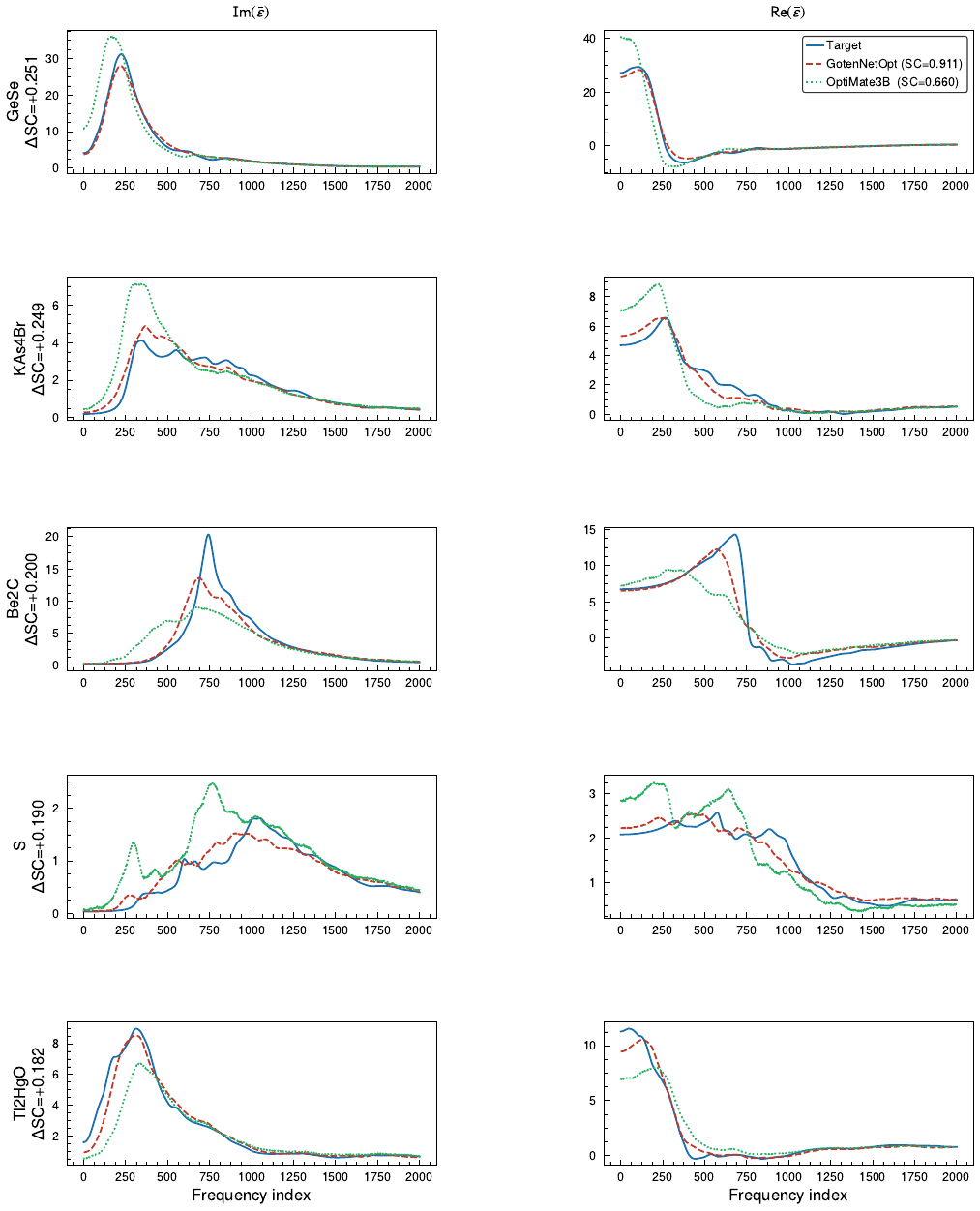}
    \caption{Predicted of \(\text{GotenNet}_\text{Opt}\) (dashed, red) and OptiMate3B (dashed, green) vs.\ target (solid, blue) dielectric
        spectra for the top 5 wins of \(\text{GotenNet}_\text{Opt}\) over OptiMate3B as measured by \(\Delta\)SC between the two models on the RPA test set.}
    \label{fig:top5_gotenNetWins}
\end{figure*}

\paragraph{Top 5 GotenNet wins as meaured by \(\Delta\)SC} Figure~\ref{fig:top5_gotenNetWins} shows the predicted and target dielectric spectra for the
five samples where \(\text{GotenNet}_\text{Opt}\) most outperforms OptiMate3B as measured by \(\Delta\)SC between the two models on the RPA test set.

\begin{figure*}[t]
    \centering
    \includegraphics[width=\textwidth]{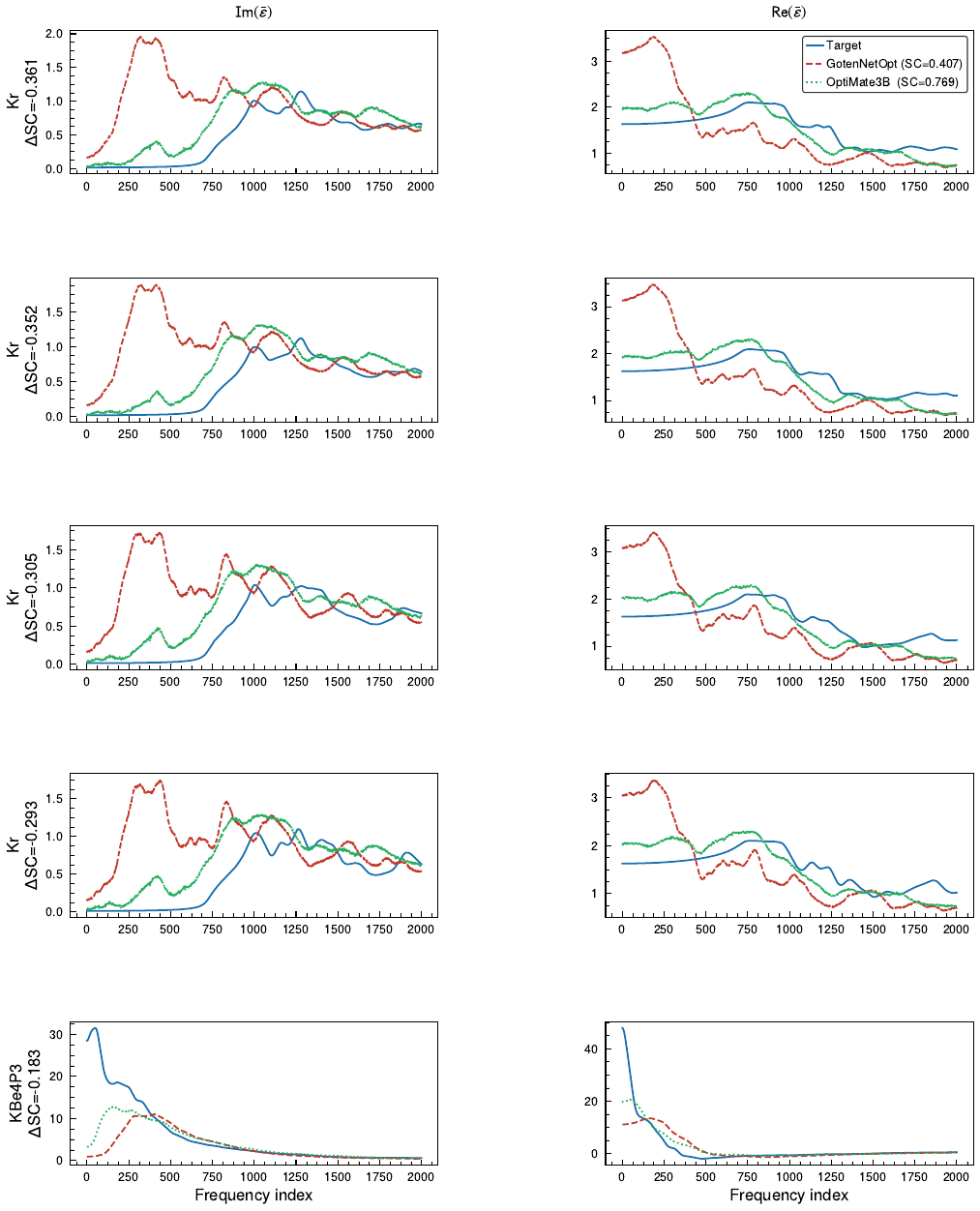}
    \caption{Predicted spectra of \(\text{GotenNet}_\text{Opt}\) (dashed, red) and OptiMate3B (dashed, green) vs.\ target (solid, blue) dielectric
        spectra for the top 5 wins of OptiMate3B over \(\text{GotenNet}_\text{Opt}\) as measured by \(\Delta\)SC on the RPA test set.}
    \label{fig:top5_optimateWins}
\end{figure*}

\paragraph{Top 5 OptiMate3B wins as measured by \(\Delta\)SC.} Figure~\ref{fig:top5_optimateWins} shows the predicted and target dielectric spectra for the
five samples where OptiMate3B most outperforms \(\text{GotenNet}_\text{Opt}\) as measured by \(\Delta\)SC.

\begin{figure*}[t]
    \centering
    \includegraphics[width=\textwidth]{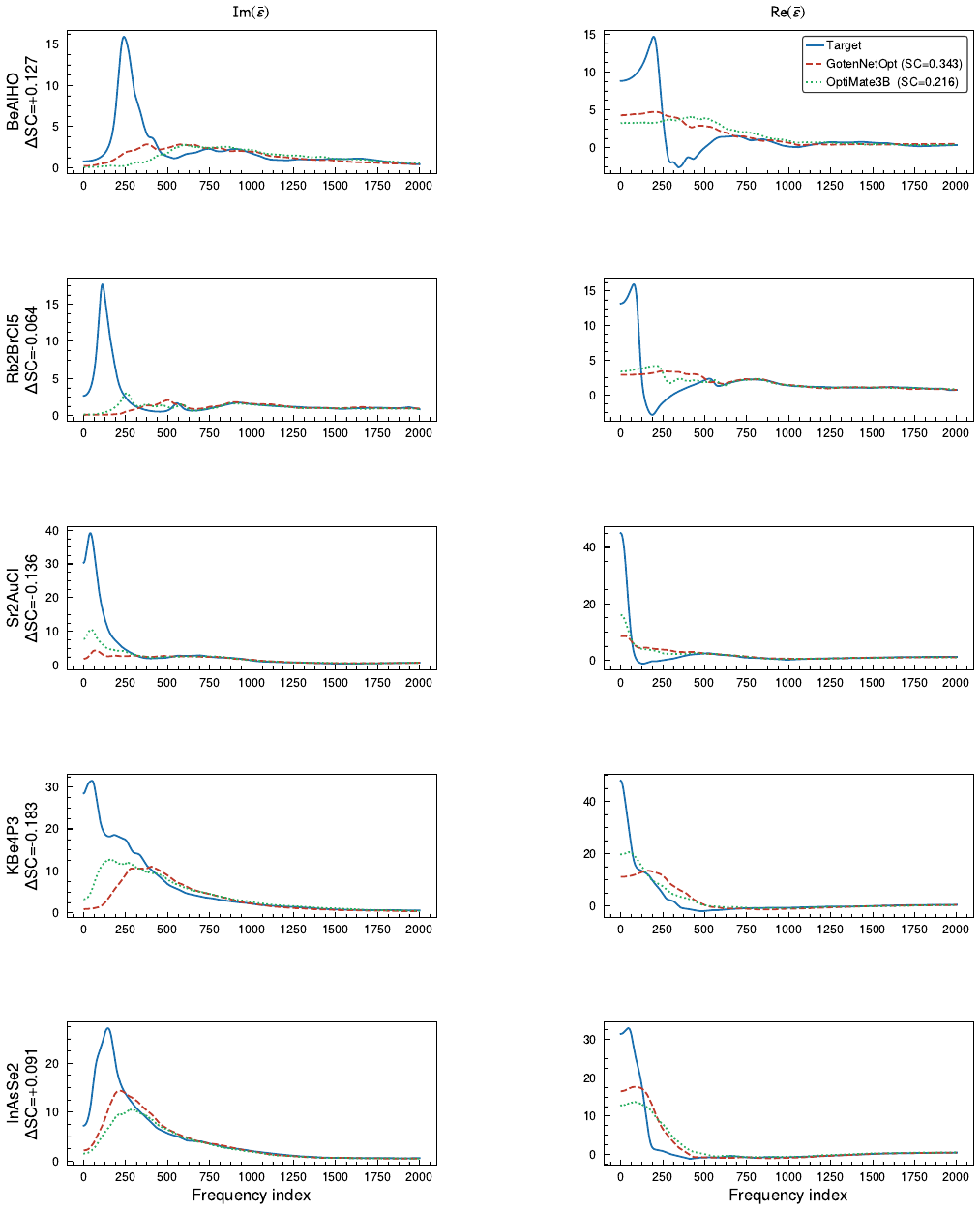}
    \caption{Predicted spectra of \(\text{GotenNet}_\text{Opt}\) (dashed, red) and OptiMate3B (dashed, green) vs.\ target (solid, blue) for the 5 samples with the lowest average SC across both models (SC\,\(<\,0.80\)) on the RPA test set, illustrating failure cases shared by both models.}
    \label{fig:top5_bothFail}
\end{figure*}

\paragraph{Top 5 samples where both models fail.} Figure~\ref{fig:top5_bothFail} shows the predicted and target dielectric spectra for the five
samples with the lowest average SC across both models, i.e.\ the cases where both models
perform worst.

\begin{figure*}[t]
    \centering
    \includegraphics[width=\textwidth]{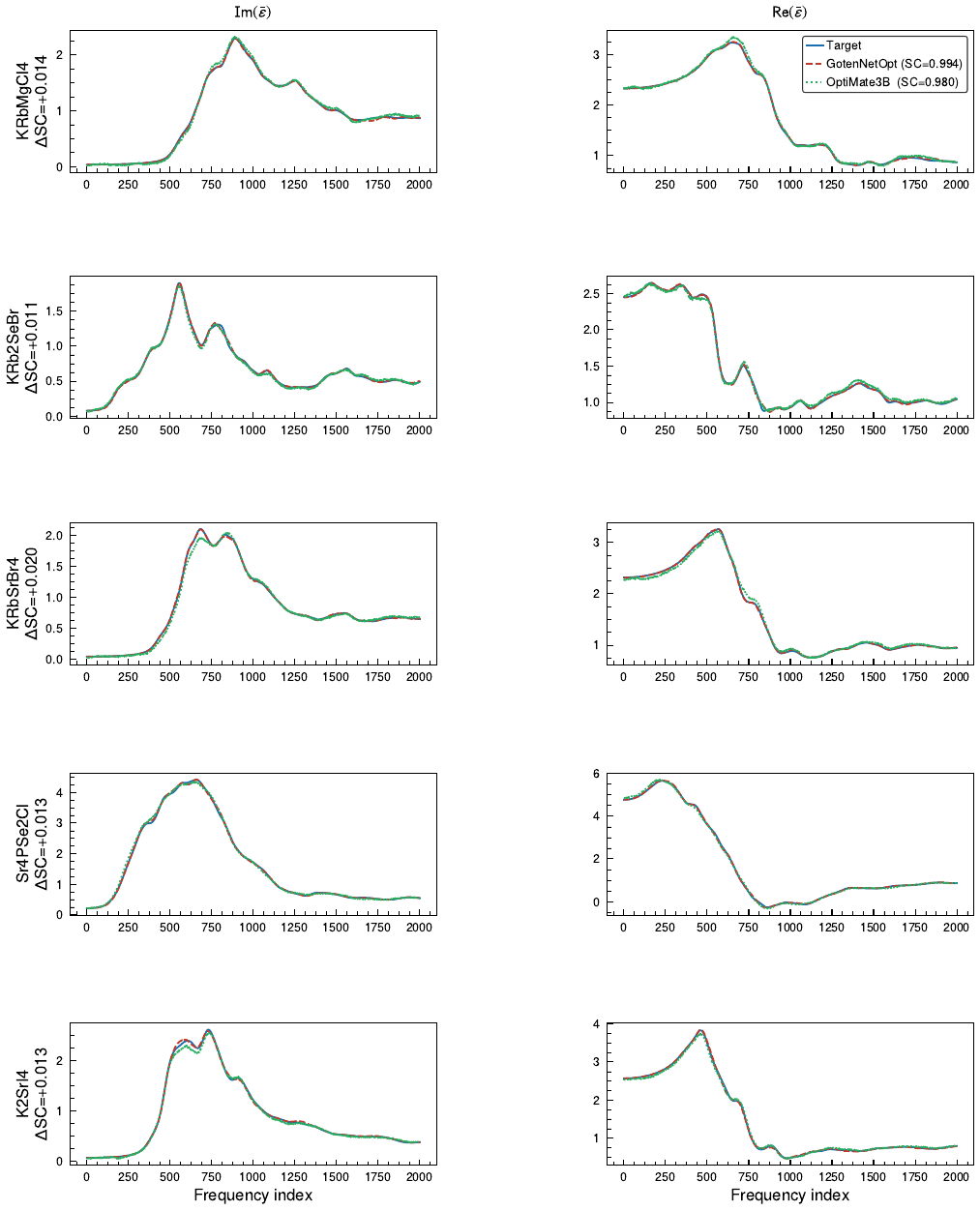}
    \caption{Predicted spectra of \(\text{GotenNet}_\text{Opt}\) (dashed, red) and OptiMate3B (dashed, green) vs.\ target (solid, blue) for the 5 samples with the highest average SC across both models (SC\,\(>\,0.95\)) on the RPA test set, illustrating representative cases where both models achieve high fidelity.}
    \label{fig:top5_bothGood}
\end{figure*}

\paragraph{Top 5 samples where both models succeed.} Figure~\ref{fig:top5_bothGood} shows the predicted and target dielectric spectra for the five
samples with the highest average SC across both models, i.e.\ the cases where both models
perform best.

\end{document}